\documentclass{aa}  

\pdfoutput=1

\usepackage{graphicx}
\usepackage[varg]{txfonts}

\usepackage{caption}
\usepackage{subcaption}

\usepackage{natbib}
\usepackage{xcolor}
\usepackage{amsmath}

\begin{document}

    \title{Sizes of protoplanetary discs after star-disc encounters}
    \subtitle{}
    \author{Andreas Breslau
       \and Manuel Steinhausen
       \and Kirsten Vincke
       \and Susanne Pfalzner
    }

\institute{Max-Planck-Institut f\"ur Radioastronomie, Auf dem H\"ugel 69, 53121 Bonn, Germany\\
    \email{abreslau@mpifr.de}
}

\date{ }

\abstract{
Most stars do not form in isolation, but as part of a star cluster or association.
These young stars are initially surrounded by protoplanetary discs.
In these cluster environments tidal interactions with other cluster members can alter the disc properties.
Besides the disc frequency, its mass, angular momentum, and energy,
in particular the disc's size is prone to being changed by a passing star.
So far the change in disc size was only investigated for a small number of very specific encounters.
Several studies investigated the effect of the cluster environment
on the sizes of planetary systems, like our own solar system,
based on a generalisation of information from this limited sample.
We performed numerical simulations covering the wide parameter space typical for young star clusters,
to test the validity of this approach.
Here the sizes of discs after encounters are presented,
based on a size definition which is comparable to that one used in observational studies.
We find that, except for encounters between equal-mass stars, the usually applied estimates are insufficient.
They tend to severely overestimate the remaining disc size.
We show that the disc size after an encounter can be described by a relatively simple dependence
on the periastron distance and the mass ratio of the encounter partners.
This knowledge allows, for example, to pin down the types of encounter possibly responsible
for the structure of today's solar system.
}


\maketitle

\graphicspath{{graphics/}}

\section{Introduction}
\label{sec:intro}

Stars form through the gravitational collapse of dense cores within molecular clouds.
Due to the angular momentum conservation during the collapse, most young stars are initially surrounded by a disc consisting of gas and dust.
These protoplanetary discs are usually treated as unperturbed by their surrounding.
However, as most stars do not form in isolation but in star cluster environments ({{Lada} \& {Lada}} 2003; {{Porras} {et~al.}} 2003)
discs are prone to be affected by external processes like photoevaporation from massive stars (e.g. {{Johnstone} {et~al.}} 1998; {{Scally} \& {Clarke}} 2001)
or gravitational interactions with other cluster members (e.g. {{Moeckel} \& {Bally}} 2006; {{Craig} \& {Krumholz}} 2013).
In this paper we only concentrate on the effect of the latter.

The consequences of gravitational interactions - or encounters - on disc parameters, like the disc's mass,
angular momentum and energy, have been intensively investigated analytically and numerically in the past
(e.g. {{Clarke} \& {Pringle}} 1993; {Ostriker} 1994; {Heller} 1995; {{Hall} {et~al.}} 1996; {{Kobayashi} \& {Ida}} 2001; {Pfalzner} 2003; {{Olczak} {et~al.}} 2006; {{Pfalzner} \& {Olczak}} 2007{a}; {{Lestrade} {et~al.}} 2011; {{Steinhausen} {et~al.}} 2012).
By contrast, the encounter induced alteration of the size of a protoplanetary disc has so far been investigated in less detail.
The reason is that only for the order of a few hundred objects, disc sizes have so far been determined (for an overview see e.g. {{Williams} \& {Cieza}} 2011 and references therein).
Deriving correlations between the disc size and the stellar environment seemed so far out of reach.
However, with the advent of ALMA in the near future the sample size is likely to increase considerably and the influence of the environment on the disc size becomes testable.

In the past, theoretical investigations have shown that encounters mainly affect the outer disc material.
{{Clarke} \& {Pringle}} (1993) investigated the effect of parabolic encounters of equal-mass stars with a periastron distance
of $1.25$ times the initial radius of the disc, $r_{\mathrm{init}} $, and varying inclinations.
They found that in the most destructive encounter (prograde, coplanar) disc material down to $\approx 0.5$ times the periastron distance, $r_{\mathrm{peri}} $, becomes unbound.
By varying the disc-size to periastron-distance ratio, {{Hall} {et~al.}} (1996) found, that in parabolic, prograde, coplanar encounters of equal mass stars
the disc inside of $0.3\,r_{\mathrm{peri}} $ is almost unperturbed.
{{Kobayashi} \& {Ida}} (2001)
focused on the encounter induced increase of eccentricity and inclination of the disc material and the size of the largely unperturbed inner part of the disc.
They found that many particles become unbound outside of $\approx 1/3$ of periastron distance after a prograde encounter
of equal-mass stars on a parabolic orbit.

Almost all previous investigations focused on a small parameter space, which was mostly restricted to encounters between equal-mass stars.
Nevertheless, the above results are often generalised as encounters truncating discs to $1/2\text{--}1/3$ of the periastron distance
(e.g. {{Brasser} {et~al.}} 2006; {{Adams} {et~al.}} 2006; {Adams} 2010; {{Jim{\'e}nez-Torres} {et~al.}} 2011; {{Malmberg} {et~al.}} 2011; {Pfalzner} 2013).
There the dependence of the disc truncation on the mass of the perturbing star and other geometrical properties of the perturber orbit are disregarded.

One typical context where the disc size after an encounter plays a key role is the formation of the solar system.
The size of the solar system, with a density drop at $\approx 30$\,AU, is attributed to an encounter of the young
solar system with another star at a periastron distance of $\approx 100$\,AU.
This distance is used to determine the typical cluster environment in which the solar system might have developed.

In addition, the disc size is a crucial parameter for investigations of the typical types of planetary systems in a given cluster environment.
Therefore, {{de Juan Ovelar} {et~al.}} (2012) tried to estimate the sizes of discs after encounters considering both, the periastron distance and the mass ratio of the stars.
They suggested two different approaches.
The first one is based on the assumption that disc material is removed in an encounter at least to the point of gravitational force equilibrium between the stars at the time of periastron passage.
In the other approach they suggested a transformation of the results for the disc-mass loss in encounters by {{Olczak} {et~al.}} (2006) into a reduction of the disc size.
Both approaches assume that disc size change and the removal of disc material are strongly correlated.

However, already {Hall} (1997) showed that during an encounter disc material can be moved inward due to loss of angular momentum.
That way, the disc size can be reduced even when no mass is lost.
More recently, it was demonstrated that a $3\text{--}5\,\%$ loss of disc angular momentum is common in ONC-like star clusters and
this holds not only for the central high-density regions but even at the outskirts of clusters ({{Pfalzner} \& {Olczak}} 2007{b}).
Therefore it is very likely, that changes of the disc size are a more common effect in clusters than the removal of disc material.

Using an extended data base of star-disc encounters we show here that mass loss based approaches are not really suitable to determine the resulting disc size.
After a short description of our numerical method in Sect.~\ref{sec:method}, a disc-size definition is given, which is comparable to observational size determinations.
In Sect.~\ref{sec:results} we present the numerical results as well as a simple fit formula.
We compare our results to previous work in Sect.~\ref{sec:comparison} and give a brief summary and conclusion in Sect.~\ref{sec:summary}.
 
\section{Method}
\label{sec:method}

\subsection{Numerical method}

We performed numerical simulations to investigate the effect of stellar encounters on the size of protoplanetary discs.
The mass of a protoplanetary disc is usually much lower than the mass of its host star
(on average \mbox{$M_{\mathrm{disc}} \approx 0.01 M_{\mathrm{star}}$}, e.g. {{Andrews} {et~al.}} 2013).
Because of the resulting low density, especially in the here relevant outer parts of the disc, and the also relatively low temperature,
the viscous timescale is long enough that the discs survive at least some million years (e.g. {{Haisch} {et~al.}} 2001).
Since this is typically much longer than the timescale of encounters,
we can mostly neglect hydrodynamical effects as well as self-gravity within the disc.
The exception are very close penetrating encounters, which we exclude anyway (see below).
We also disregard photoevaporation from the stars as well as radiation transport within the disc.
Thus, we only consider gravitational forces from and onto the stars.

For pure gravitational numerical investigations, particle based simulations are most suitable.
Since we investigate low-mass discs and neglect the self-gravity, we can even go one step further and use mass-less tracer particles.
This has the advantage that the particle distribution can be chosen independently from the mass distribution within the disc.
To obtain a higher spatial resolution at the outskirt of the disc, we use a constant particle surface density ($\Sigma(r) = \mathrm{const.}$).
The particle masses are associated to the tracer particles in the diagnostic step, making it also possible to investigate the effect of different initial mass distributions
with one single simulation ({{Steinhausen} {et~al.}} 2012).

In dimensionless units, our discs have an initial outer radius of $r_{\mathrm{init}}  = 1$ around a star of mass $M_{\mathrm{1}} = 1$.
Because the test particles are only affected by gravitational forces, they initially follow keplerian orbits around their host star.
We assume that all particles are initially on circular orbits (eccentricity $e = 0$).

Even though hydrodynamical forces are neglected, we set up our discs with a certain thickness
for consistency with previous work (e.g. {{Pfalzner} {et~al.}} 2005{b}; {{Olczak} {et~al.}} 2006; {{Pfalzner} \& {Olczak}} 2007{a}; {{Steinhausen} {et~al.}} 2012).
Thus the initial particle distribution in the disc is given by
\begin{align}
\rho(r,z) \propto \Sigma(r) \exp{\left( - \frac{z^2}{2 H^2(r)} \right)},
\end{align}
where $H(r) = 0.05\,r$ the vertical half-thickness of the disc (see also {Pringle} 1981).
Since this results in inclinations $i \lesssim 5^{\circ}$ for nearly all particles the disc can still be regarded as thin.

A standard method when investigating star-disc encounters is to exclude an inner disc region to avoid unnecessary small time steps.
Here, this region extends to $r_{\mathrm{hole}} = 0.1\,r_{\mathrm{init}} $.
Therefore, encounters resulting in very small final discs with $r_{\mathrm{final}}  \lesssim 0.2\,r_{\mathrm{init}} $ are excluded from further diagnostics,
as they might be influenced by the missing disc mass within the hole.
Particles which approach one of the stars closer than $0.1\,r_{\mathrm{hole}}$ are treated as accreted.
As a compromise between spatial resolution and computation time needed for the parameter study, our discs are modelled with $10\,000$~particles
(see also {{Kobayashi} \& {Ida}} 2001; {Pfalzner} 2003; {{Steinhausen} {et~al.}} 2012).

Simulations were performed for varying perturber-mass to host-mass ratios ${m_{\mathrm{12}}}  = M_{\mathrm{2}}/M_{\mathrm{1}}$ and periastron-distances $r_{\mathrm{peri}} $.
The trajectories of the particles were integrated by using a Runge-Kutta Cash-Karp scheme with adaptive time step size.

As parameter space for our study, we have chosen the mass-ratio and periastron-distance range
typical for young dense clusters in the solar neighbourhood.
A prime example is the Orion Nebula Cluster (ONC), for which the effect of encounters on the protoplanetary discs was already investigated by {{Pfalzner} \& {Olczak}} (2007{b}).
The mass range in the ONC is given by the mass ratio of the stars at helium burning limit, $M \approx 0.08\,\mbox{M$_{\odot}$} $,
and the most massive system $\theta^1$~Ori~C with $M \approx 40\,\mbox{M$_{\odot}$} $.
Previous studies (e.g. {{Pfalzner} {et~al.}} 2005{b}; {{Steinhausen} {et~al.}} 2012) have shown that encounters with very low mass ratios (${m_{\mathrm{12}}}  \lesssim 0.1$) only have
an effect on the discs for very small periastron distance ($r_{\mathrm{peri}}  \lesssim r_{\mathrm{init}} $).
Additionally, hydrodynamical effects may play a role in these encounters, and the mass of the disc is not much smaller than the mass of the perturber,
which would make the application of the here used method questionable.
Therefore, we treat only the range ${m_{\mathrm{12}}}  = 0.3\text{--}500$.

We restrict the periastron distances to the range where the discs are perturbed significantly but not fully destroyed.
Cases where the resulting disc sizes are $\lesssim 0.2\,r_{\mathrm{init}} $ are excluded from further diagnostics, as they might be influenced by the inner cut-off of the initial discs.
For an equal mass perturber the distances range, for example, from $0.7\,r_{\mathrm{init}} $ to $7\,r_{\mathrm{init}} $.
The actual periastron distances depend on the mass ratio and can be found in Table~\ref{tab:sizes} in the Online Material.

In this work, we only consider prograde, coplanar and parabolic ($e = 1$) encounters as they are the most destructive ones.
Therefore our resulting disc sizes represent the lower limits for the disc sizes after encounters on inclined and/or hyperbolic orbits
(e.g. {{Clarke} \& {Pringle}} 1993; {{Pfalzner} {et~al.}} 2005{b}).
For this parameter space several hundreds of simulations were performed.

At the onset of our simulations the perturber is placed at a distance $d_{\mathrm{init}}$, where its gravitational force onto the closest disc particle is only $1\%$
relative to the force from the disc-hosting star. This condition is satisfied, when
\begin{align}
d_{\mathrm{init}} \gtrsim 10\,r_{\mathrm{init}}  \sqrt{m_{12}} +r_{\mathrm{init}} .
\end{align}
We end our simulations, when the perturber has again at least a distance of $d_{\mathrm{init}}$ to the disc-hosting star
or a particle at $r_{\mathrm{init}} $ fulfilled at least $1.5$ orbits around the host star after periastron passage.

We investigate here the simplified case of only one star being surrounded by a disc to limit computational costs.
This approach is valid, as long as only small fractions of the disc material are captured by the perturbing star ({{Pfalzner} {et~al.}} 2005{a}).
For a realistic modelling of close disc-disc encounters hydrodynamical effects would also have to be considered.

\subsection{Size determination}
\label{sec:method_size}

During the diagnostic step masses are attributed to the tracer particles so that the initial disc matches
\begin{align}
\rho(r) = \rho_0 r^{-p}\label{eq:densitydistribution},
\end{align}
where $\rho_0$ is the mass density in the equatorial plane at the inner rim ($r = 0.1\,r_{\mathrm{disc}} $) and $p$ the parameter for the slope of the mass density distribution
(see also {{Steinhausen} {et~al.}} 2012).

A problem in the determination of the size of a protoplanetary disc after an encounter is, that there is no general definition of a disc size at hand.
Taking observations as a guide is only of limited help, as even in the unperturbed case there are several methods used for disc size determination.
One common method is to fit the observed SED to disc models with radial density and temperature profiles which follow mostly a truncated power law
(e.g. {{Andrews} \& {Williams}} 2007).
The radius of truncation of the density is then taken as size.
In case of resolved images, the size corresponds often to the radius where the luminosity falls below a certain limit
(e.g. {{Vicente} \& {Alves}} 2005; {O'dell} 1998).
This drop in luminosity is equivalent to a drop in the disc's surface density.
Since both methods rely somehow on a drop in the disc's volume or surface density we decided to mimic this methods
by defining our disc size as the point of strongest contrast in the surface density.

In the determination of the disc's surface density after an encounter from the simulations, a problem arises from the particles on highly eccentric orbits.
While in the viscosity-free case the semi-major axis $a$ and eccentricity $e$ of the particles do not change any more after the end of the encounter,
their radial distances to the disc-hosting star change with time.
Therefore, global quantities determined from a snapshot of the particle distribution at a certain time would not necessarily be representative.

For better comparison of our disc size values with observational data, we use therefore a temporally averaged surface density distribution for the disc size determination.
This can be obtained directly from the data of the last time step by applying the radial probability functions of the orbits defined by $a$ and $e$.
The sum of the radial probability functions of all particles yields the time-averaged {\em particle} surface density distribution.
When weighting the radial probability functions in this sum with the previously associated particle masses according to Eq.~\eqref{eq:densitydistribution}
one obtains the {\em mass} surface density distribution.
For our diagnostics we used particle masses according to an initially $\propto r^{-1}$ ($p=1$) mass surface density distribution.

The disc size is then determined as the point of strongest mass surface density gradient.
Here, the steep inner part of the initial $r^{-1}$ mass distribution is excluded.
Due to the statistical nature of our data, the surface density distributions have to be smoothed before the size determination algorithm is applied.

\begin{figure}
  \centering
  \begin{subfigure}[t]{\hsize}
    \begin{minipage}[t]{0.03\hsize}
      \vspace{0pt}
      \caption{}\label{fig:decorrelated}
    \end{minipage}
    \begin{minipage}[t]{0.96\hsize}
      \vspace{0pt}
      \includegraphics[width=\textwidth]{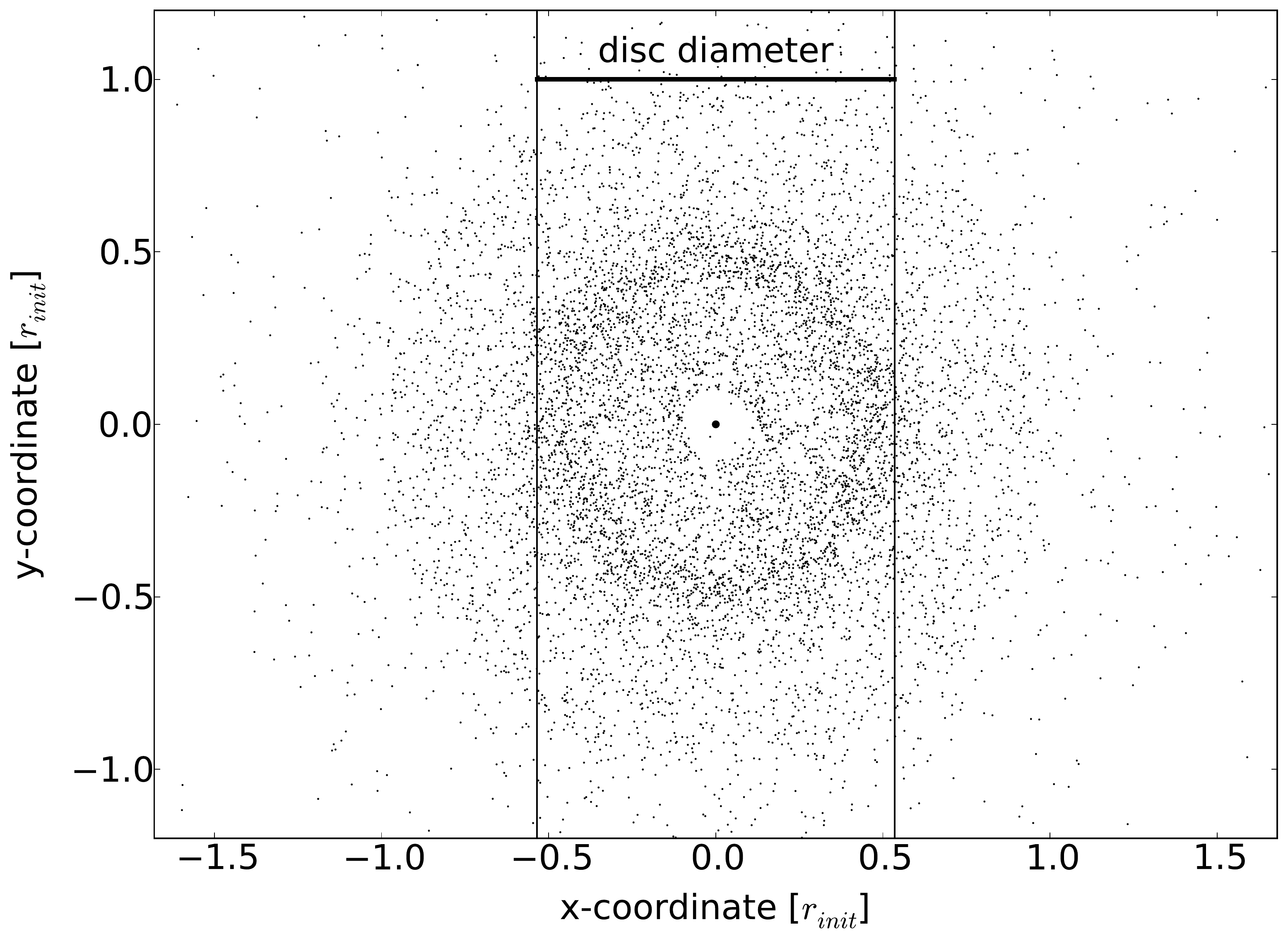}
    \end{minipage}
  \end{subfigure}
  \begin{subfigure}[t]{\hsize}
    \begin{minipage}[t]{0.03\hsize}
      \vspace{0pt}
      \caption{}\label{fig:disc_size}
    \end{minipage}
    \hfill
    \begin{minipage}[t]{0.96\hsize}
      \vspace{0pt}
      \includegraphics[width=\textwidth]{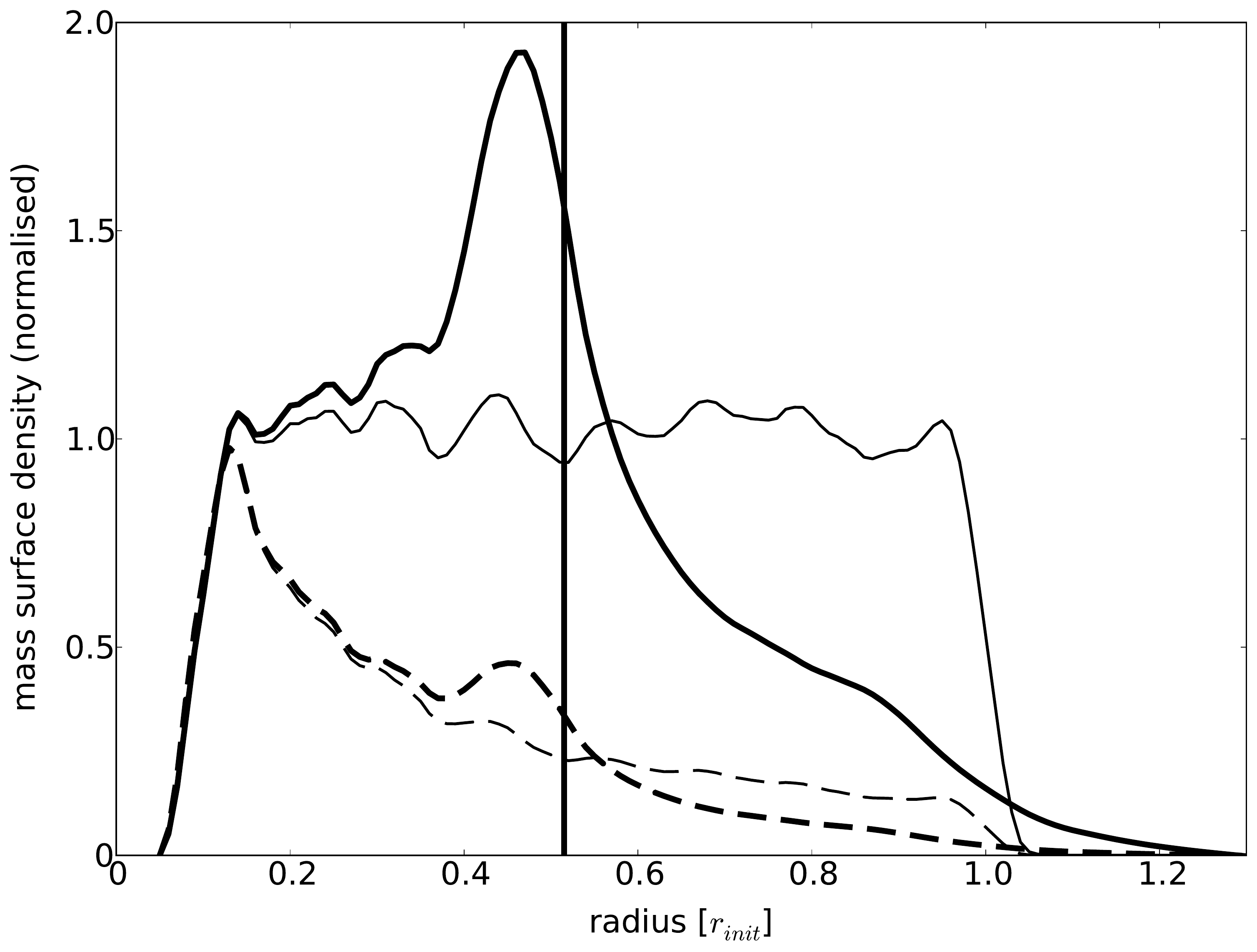}
    \end{minipage}
  \end{subfigure}
  \caption{{\bf a)} Face on view of a disc after an encounter with ${m_{\mathrm{12}}}  = 1$ and $r_{\mathrm{peri}}  = 2\,r_{\mathrm{init}} $.
    For clarity, the particle angles have been decorrelated to destroy the spiral arms.
    The vertical lines mark the disc diameter ($= 2 r_{\mathrm{disc}} $) as obtained with our disc size definition.\newline
    {\bf b)} Corresponding initial (thin) and final (thick) particle (solid) and mass (dashed) surface density distribution (all lines are smoothed).
    The vertical line shows the final disc size as obtained with our disc size definition.}
  \label{fig:size}
\end{figure}

Figure~\ref{fig:decorrelated} shows a disc perturbed by a ${m_{\mathrm{12}}}  = 1$ perturber on a parabolic orbit with $r_{\mathrm{peri}}  = 2\,r_{\mathrm{init}} $.
Figure~\ref{fig:disc_size} shows the corresponding initial (thin) and final (thick) time averaged surface density distributions
for in initially constant particle (solid) and $\propto r^{-1}$ mass (dashed) distribution.
The vertical line shows the final disc size as obtained with our disc size definition.
It can be seen, that the vertical line matches the points of steepest gradient in both final density distributions.

To determine the statistical deviations of our disc sizes we performed $\approx 100$ simulations of one encounter
with different random seeds for the initial particle distribution.
Because smaller radial bins for the surface density computation require a longer smoothing length (as mentioned above) and vice versa,
we found a bin width of $0.01\,r_{\mathrm{init}} $ as a good compromise between resolution and smoothing.
The statistical deviations are then also on the order of $0.01\,r_{\mathrm{init}} $.
 
\section{Results}
\label{sec:results}

\begin{figure}
  \centering
  \includegraphics[width=\hsize]{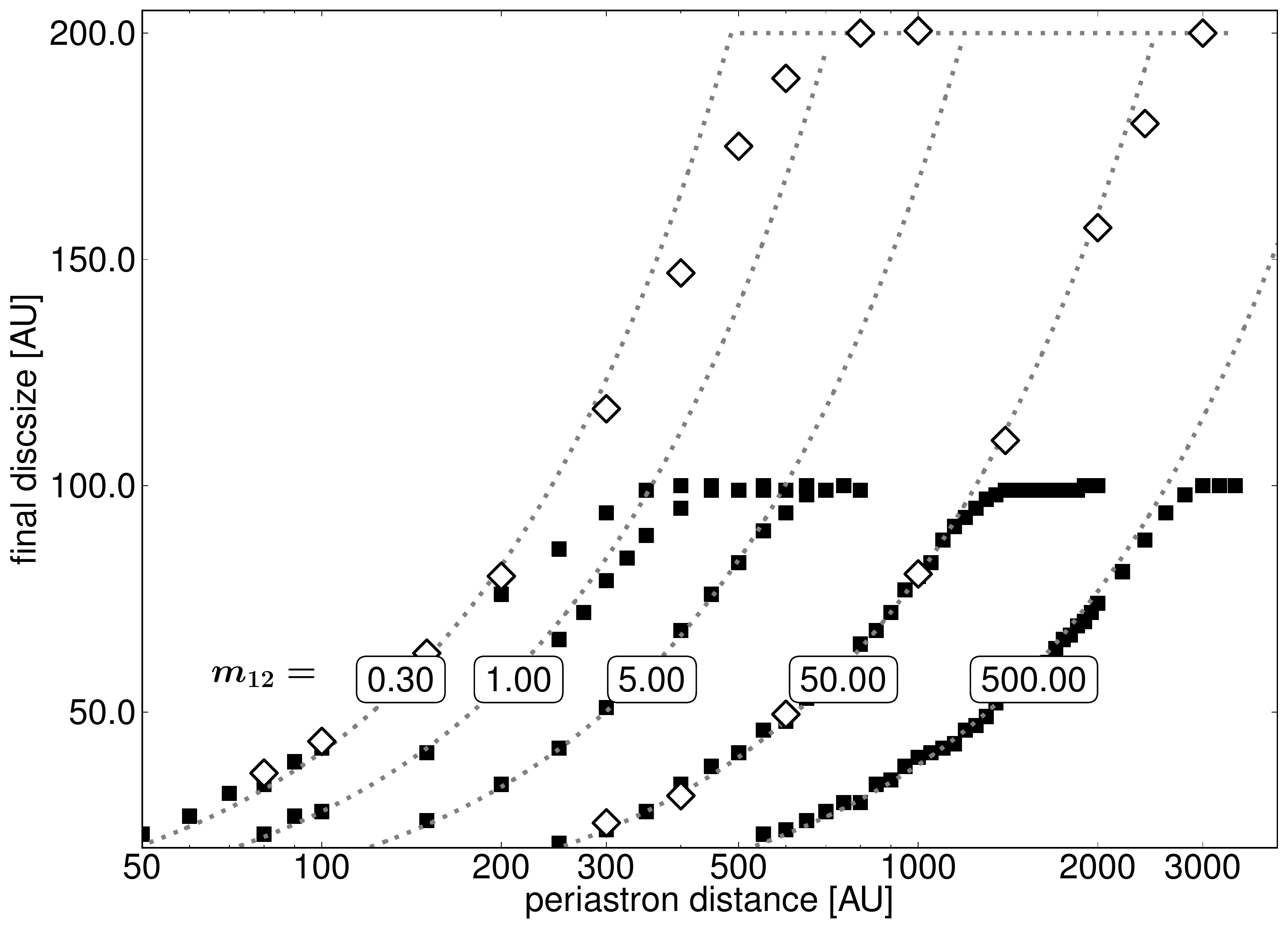}
  \caption{Final disc sizes versus perturber periastron for some perturber mass ratios (values shown by the numbers at the lines)
        from our simulations (black squares for initially $100$\,AU discs and diamonds for initially $200$\,AU discs)
        compared to our fit formula~(Eq.~\ref{eq:fitformula}, grey, dotted lines).}\label{fig:disc_sizes}
\end{figure}

Figure~\ref{fig:disc_sizes} gives an overview of the results of our parameter study.
(The actual values can also be found in Table~\ref{tab:sizes} in the Online Material.)
For clarity and better comparability to observations we present the numerical results in absolute values,
assuming an initial disc size of $100$\,AU (black squares).
For comparison, the diamonds represent some exemplary results for discs with an initial size of $200$\,AU.
As long as an encounter changes the sizes of the discs with both initial sizes significantly the resulting sizes are approximately the same.
Since the discs do not become bigger\footnote{according to our size definition} in our simulations, the final sizes are limited by the initial disc sizes.
Therefore, the final disc sizes of initially $100$\,AU and $200$\,AU discs have to deviate after encounters, where the final size of the $200$\,AU disc is $\gtrsim 100$\,AU.
For example an encounter with a mass ratio of ${m_{\mathrm{12}}}  = 0.3$ and a periastron distance of $500$\,AU
reduces the size of an initially $200$\,AU disc to $\approx 150$\,AU while the initially $100$\,AU disc is unchanged.

There exists a relatively simple dependence of the disc size on the mass ratio between the encountering and the disc-bearing star (${m_{\mathrm{12}}} $)
and the periastron distance ($r_{\mathrm{peri}} $) which can be described by a fit formula of the form
\begin{align}
  r_{\mathrm{final}}  = 0.28 \cdot {m_{\mathrm{12}}} ^{-0.32} \cdot r_{\mathrm{peri}}  \label{eq:fitformula}.
\end{align}
Again, the final disc sizes are limited by the initial disc sizes, as the discs do not grow.
In the plot the fit function is limited by the initial size of the initially $200$\,AU discs.
Close to the sharp upper cut off of the fit function, the numerical values deviate slightly from the fit since the curves bend smoothly.

The deviations of the final disc sizes obtained with Eq.~\eqref{eq:fitformula} from the results of the simulations are $\lesssim 5\,\%$ of the initial disc size
(i.e. $5$\,AU for an initially $100$\,AU disc) for mass ratios ${m_{\mathrm{12}}}  = 5\text{--}500$
and $\lesssim 10\,\%$ for mass ratios ${m_{\mathrm{12}}}  = 0.3\text{--}5$.
These deviations are still on the order of the uncertainty of the size definition.
The validity of our fit formula is also restricted to periastra where $r_{\mathrm{final}}  > 0.2\,r_{\mathrm{init}} $.

\section{Discussion}
\label{sec:discussion}

Some approximations have been made in the model described above.
First, we treated the discs by pure N-body methods while neglecting viscous forces.
Viscous forces only play a role in the central areas of the discs.
For typical viscosity values the effected area is the region within $ \lesssim 0.2\,r_{\mathrm{init}} $.
Only in the most violent interactions the disc size is reduced to such low values.
For these cases viscous forces would have to be in principle included.
As can be seen in Table~\ref{tab:sizes}, this is, for example, the case for equal-mass stars in penetrating encounters closer than $ 0.7\,r_{\mathrm{init}} $.
For typical disc sizes on the order of some $100$\,AU (e.g. {{Bally} {et~al.}} 2000; {{Andrews} {et~al.}} 2009), such encounters are relatively rare in ONC-Type clusters ({{Olczak} {et~al.}} 2006).
In addition, for such destructive encounters the material that remains bound is usually $<20\,\%$ of its initial mass
and its structure does not resemble a disc as such.
In these cases the definition of a disc size is anyway highly questionable.

An additional approximation is that the case where only one of the stars was surrounded by a disc was considered.
This has been done for simplicity of description.
In principle, for disc-disc encounters the sizes could be larger because disc material could be transferred from the perturbing star
to the disc hosting star and replenish the disc.
{{Pfalzner} {et~al.}} (2005{a}) showed that in case both stars are surrounded by discs,
mostly an additive approach can be used as usually captured material is deposited close to the star.
In addition, the amount of captured mass is very small compared to the disc mass as such.
Therefore in most cases the final disc size is little influenced by captured material.
The exception is again the case where the encounter is very violent and the remnant disc mass very low.

Furthermore, the outcome of a disc-disc encounter could be influenced by the dependence of the disc mass and size on the stellar mass.
Observations give no conclusive answer to this question.
If the disc size scales in some way with the stellar mass, the situation can occur that the perturber has a much higher mass,
and therefore has a bigger and possibly more massive disc (see e.g. {{Andrews} {et~al.}} 2013).
In this case the amount of material captured by the primary could be comparable to or even higher than that of its remaining disc.

We have chosen the point of highest surface density gradient as definition of the disc size (see Sect.~\ref{sec:method_size}).
However, after an encounter the region where the disc's surface density drops significantly spans some range (see Fig.~\ref{fig:disc_size}).
The inner boundary of this region would correspond better to the disc sizes obtained by SED fits,
where the size is usually defined as the point, where the surface density transits from a power-law to something steeper.
If we had defined our disc sizes as the inner boundary of the density-drop region,
the results would be smaller than with our actual definition.
Conversely, if defining the size as the point, where the surface density falls below a certain threshold (compare Sect.~\ref{sec:method_size}),
the sizes would be somewhat bigger than our results.
However, results obtained with each of these definitions would usually not differ from our results by more than $\approx 0.1\,r_{\mathrm{init}} $.
Our definition is therefore a robust mean value between several possible disc size definitions for perturbed discs.

Since the problem of star disc encounters scales with the periastron ratio, one would expect that the results can be normalised to this ratio $r_{\mathrm{peri}} /r_{\mathrm{init}} $.
This is indeed the case, but for more intuitive comparison with observations we have chosen the absolute presentation of the results.
The actual values in Table~\ref{tab:sizes} in the Online Material are normalised to the periastron ratio.

The differences between the simulation results and the fit formula increase for decreasing mass ratios, becoming significant ($\gtrsim 5\%$) for ${m_{\mathrm{12}}}  \lesssim 5$.
This is because in the close encounters needed to change the disc size for low mass ratios the redistribution of the disc material is non-linear.
Therefore the fit formula with linear dependence on the periastron distance can describe this effect only approximately.

It is also important to note, that in contrast to close, penetrating encounters,
the disc size change in distant encounters is not mainly caused by a real truncation of the disc.
Mostly, the size change is dominated by the redistribution of disc material towards the host star.
This redistribution results in smaller disc sizes even when no material is lost at all.

\begin{figure}
  \centering
  \includegraphics[width=\hsize]{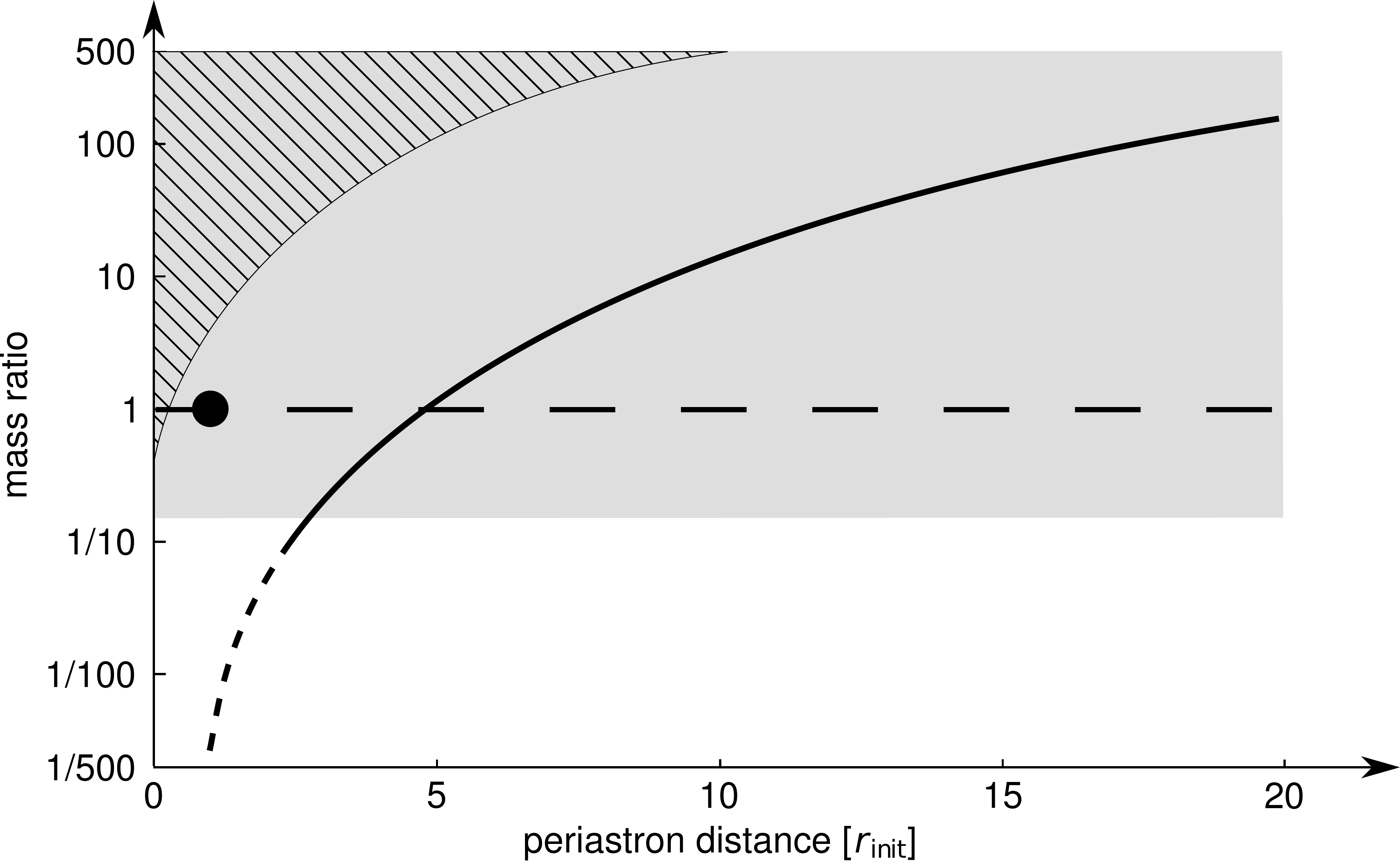}
  \caption{The parameter space of encounters occurring in a star cluster like the ONC (full height and open to the right)
    compared to the parameter space of our simulations (grey area) and the validity area of our fit formula (grey, not ruled area).
    For further details see text.}
  \label{parameter_space}
\end{figure}

Figure~\ref{parameter_space} shows where our fit formula can be applied.
The axes are chosen to cover the encounter parameter space typical for an ONC-like star cluster.
The grey area shows the parameter space covered by this study.
Right from the solid bent line, encounters have almost no effect on the mass, angular momentum or size of a disc.
Right from the dashed bent line, it can be assumed that encounters have also a negligible effect.
The horizontal, dashed line depicts the parameter space of {{Hall} {et~al.}} (1996)
while the black dot marks the simulations by {{Kobayashi} \& {Ida}} (2001).
In the black shaded area the remaining discs have sizes below $0.2\,r_{\mathrm{init}} $
or have less than $10\,\%$ of their initial particles.
Since the simulations in this area may be influenced by low resolution or the hole in our discs,
we excluded them.
The remaining grey area depicts the parameter range for which our fit formula is valid.

{{Kobayashi} \& {Ida}} (2001) developed an analytical estimate for the size of the inner disc region,
where the velocity dispersion of the disc material after an encounter is still low enough to form planets (see their Eq.~(31)).
They obtained 
\begin{align}
  r_{\mathrm{planet}} \propto \left( \frac{{m_{\mathrm{12}}}  +1}{{m_{\mathrm{12}}} ^2} \right)^{1/4} \left(\frac{r_{\mathrm{peri}} }{r_{\mathrm{init}} } \right)^{5/4}.
\end{align}
Even though the size of this planet forming region is defined differently than our disc size,
the mass dependence is similar to the ${m_{\mathrm{12}}} ^{-0.32}$ dependence of Eq.~\eqref{eq:fitformula}.
However, the periastron dependence of their analytical approximation is stronger than that of our numerical results.

\section{Comparison with previous approximations}
\label{sec:comparison}

\begin{figure}[t!]
  \centering
  \begin{subfigure}[t]{\hsize}
    \begin{minipage}[t]{0.03\hsize}
      \vspace{0pt}
      \caption{}\label{fig:cmp_one_third}
    \end{minipage}
    \hfill
    \begin{minipage}[t]{0.96\hsize}
      \vspace{0pt}
      \includegraphics[width=\textwidth]{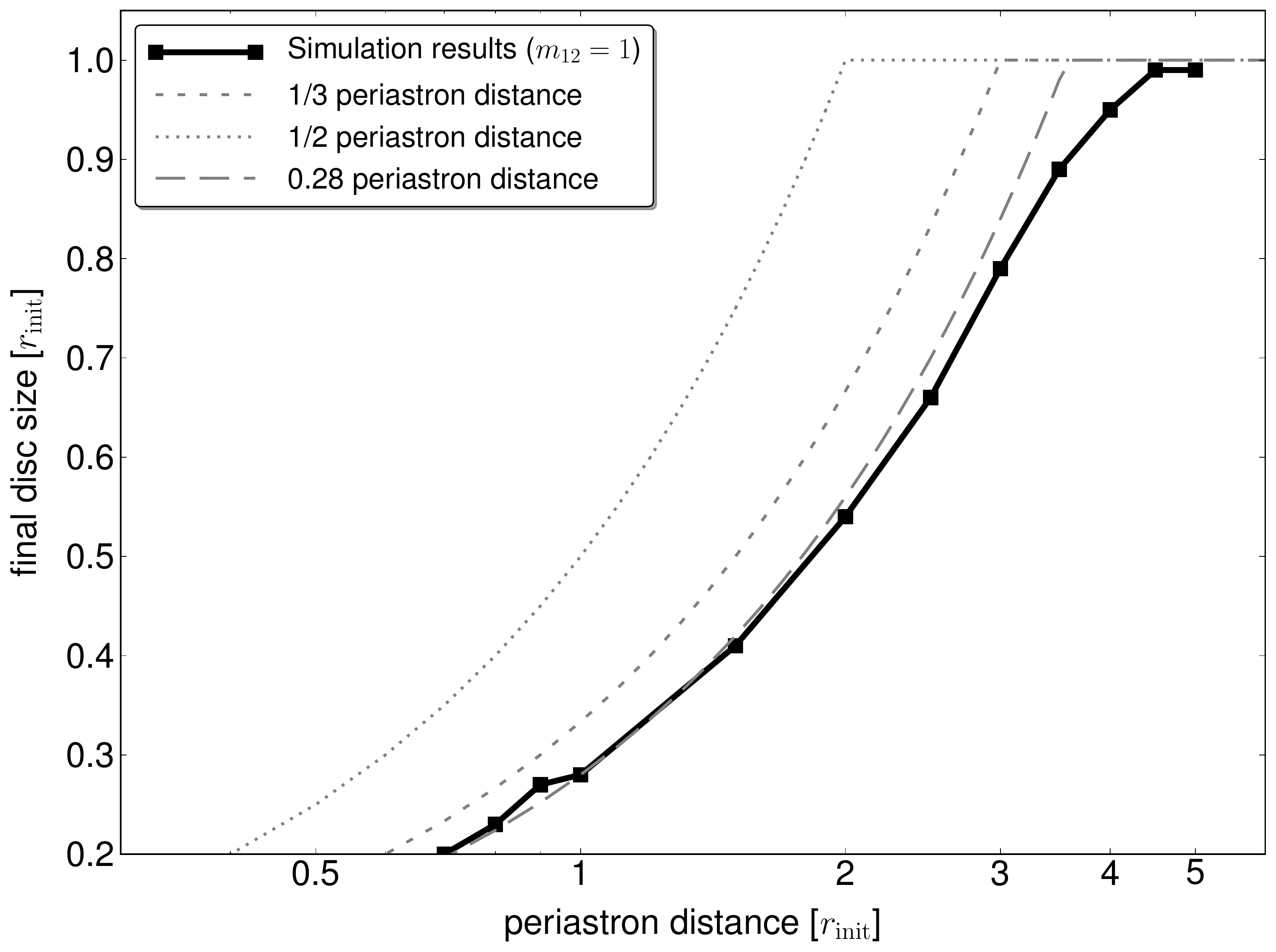}
    \end{minipage}
  \end{subfigure}
  \begin{subfigure}[t]{\hsize}
    \begin{minipage}[t]{0.03\hsize}
      \vspace{0pt}
      \caption{}\label{fig:cmp_equal_force}
    \end{minipage}
    \begin{minipage}[t]{0.96\hsize}
      \vspace{0pt}
      \includegraphics[width=\textwidth]{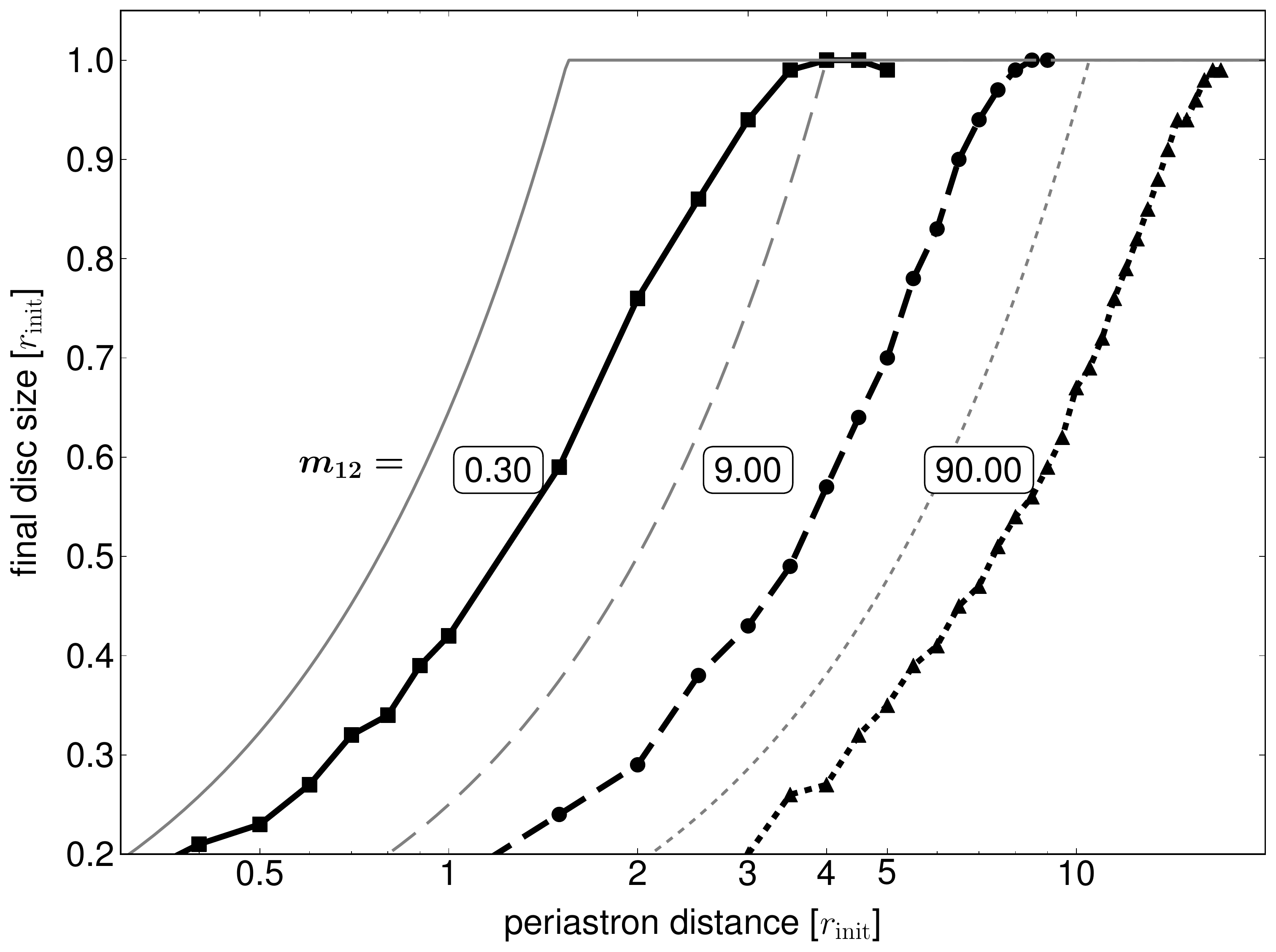}
    \end{minipage}
  \end{subfigure}
  \begin{subfigure}[t]{\hsize}
    \begin{minipage}[t]{0.03\hsize}
      \vspace{0pt}
      \caption{}\label{fig:cmp_mass_loss}
    \end{minipage}
    \begin{minipage}[t]{0.96\hsize}
      \vspace{0pt}
      \includegraphics[width=\textwidth]{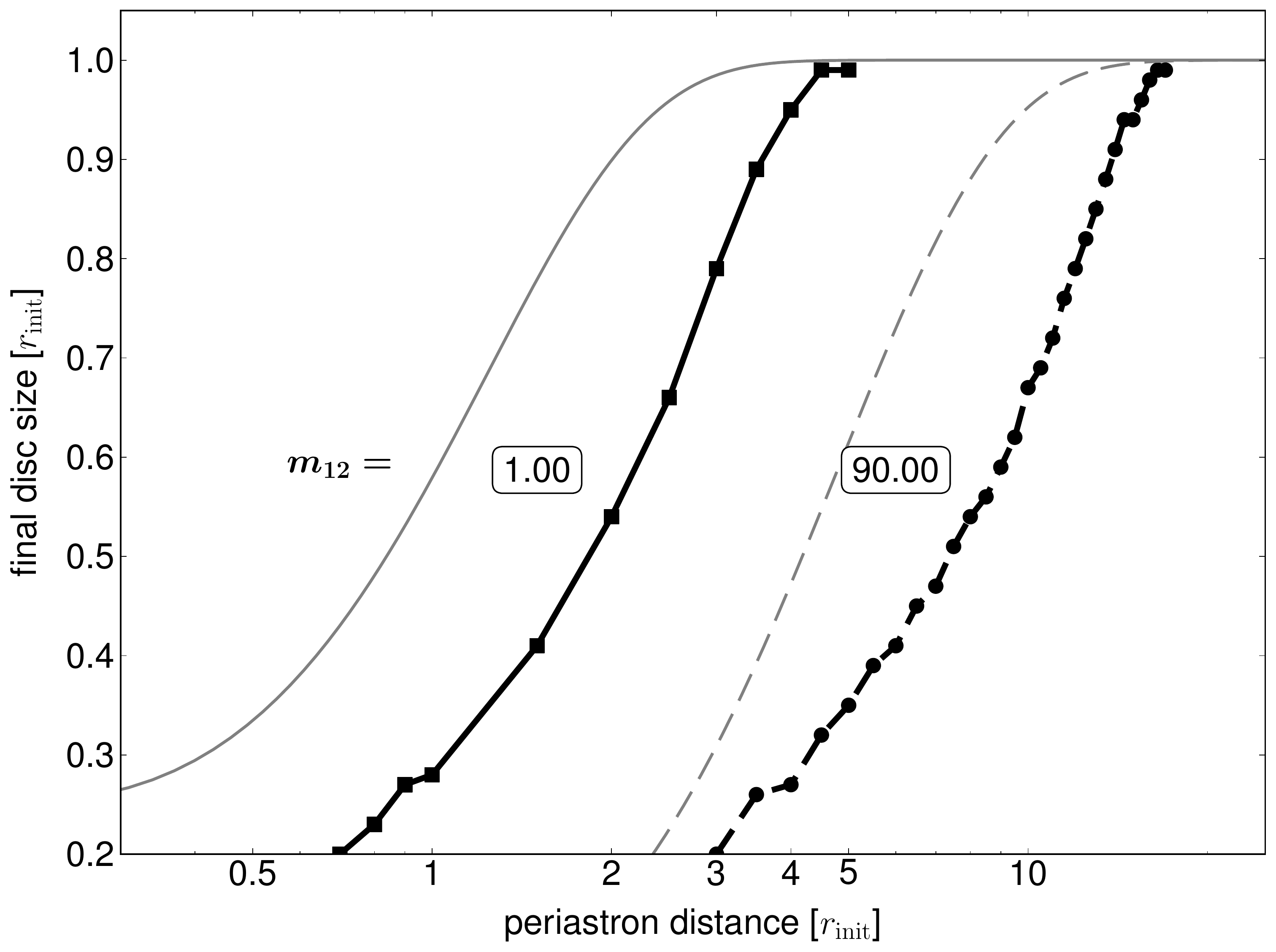}
    \end{minipage}
  \end{subfigure}
  \caption{Disc sizes from our simulations (black lines) versus perturber periastron for
    {\bf a)} an equal mass perturber compared to $1/2$ (dotted grey line) and $1/3$ (short dashed grey line) of the periastron distance
    as well as our new fit formula (long dashed grey line).
    {\bf b)} perturbers with $0.3$~(solid), $9.0$~(dashed) and $90.0$~(dotted) solar masses compared to the disc sizes obtained with Eq.~\eqref{eq:equal_force} (grey lines).
    {\bf c)} perturbers with $1.0$~(solid) and $90.0$~(dotted) solar masses compared to the disc sizes obtained with Eq.~\eqref{eq:mass_loss} (grey lines).}
  \label{fig:cmp_previous_work}
\end{figure}

Figure~\ref{fig:cmp_one_third} shows a comparison of the numerical results (solid line) to the often used approximations (see Sect.~\ref{sec:intro})
\begin{equation}
  r_{\mathrm{final}}  = \frac{1}{3} r_{\mathrm{peri}}  \mathrm{\hspace{1em} and \hspace{1em}}r_{\mathrm{final}}  = \frac{1}{2} r_{\mathrm{peri}}  \label{eq:approximations}
\end{equation}  
($1/3\,r_{\mathrm{peri}} $ - short dashed line, $1/2\, r_{\mathrm{peri}} $ - dotted line).
These approximations obviously can only be applied to encounters with $r_{\mathrm{peri}} /r_{\mathrm{final}}  < 3$ and $r_{\mathrm{peri}} /r_{\mathrm{final}}  < 2$, respectively.
Therefore in Fig.~\ref{fig:cmp_one_third} the disc sizes for wider perturber periastra are truncated at $r_{\mathrm{final}} /r_{\mathrm{disc}}  =1$.

The $1/2\,r_{\mathrm{peri}} $-approach is always much bigger than the numerical results and never a good approximation of the disc size after an encounter.
The approach of using a third of the periastron distance works when the size after the encounter lies in the range of $0.2\text{--}0.4\,r_{\mathrm{init}} $,
equivalent to penetrating or grazing encounters ($r_{\mathrm{peri}} /r_{\mathrm{init}}  = 0.6\text{--}1.2$).
This means, that we confirm the result of {{Kobayashi} \& {Ida}} (2001) for solar system forming encounters with solar-type stars.
However, a generalisation to more distant encounters common in cluster environments yields much too big disc sizes.
The difference between the simulation results and the $1/3$ approximation is up to $\gtrsim 20\,\%$,
the difference to the $1/2$ approximation up to $\approx 80\,\%$.
For equal mass stars our fit formula Eq.~\ref{eq:fitformula} simplifies to
\begin{align}
  r_{\mathrm{final}}  =  0.28 \cdot r_{\mathrm{peri}} .
\end{align}
This formula can reproduce the numerical results slightly better with deviations $\lesssim 10\%$ (see also the long dashed grey line in Fig.~\ref{fig:cmp_one_third}).
Nevertheless, a generalisation of any of these mass-ratio independent approximations to other than equal-mass encounters can obviously lead to tremendous errors.

Recently, {{de Juan Ovelar} {et~al.}} (2012) derived an upper limit of the disc size after an encounter as the distance $d_{eq}$
at which the gravitational force of the perturbing star dominates at the time of pericentre passage.
This is given by
\begin{align}
  d_{eq} = \frac{r_{\mathrm{peri}} }{1+{m_{\mathrm{12}}} ^{1/2}}. \label{eq:equal_force}
\end{align}
Figure~\ref{fig:cmp_equal_force} shows the comparison of the sizes approximated
by Eq.~\eqref{eq:equal_force} (thin, grey lines) with the simulation results (thick, black lines).
For clarity we show only the curves for three mass ratios ($0.3$~(solid), $9.0$~(dashed) and $90.0$~(dotted)).
Evidently, at least for prograde, coplanar encounters, this analytical approximation can be only regarded as upper limit,
because the actual disc sizes are typically between a factor of one and two smaller than this estimate.

In addition, {{de Juan Ovelar} {et~al.}} (2012) suggested to use the mass loss in an encounter to estimate the resulting disc size.
Using the mass-loss expression by {{Olczak} {et~al.}} (2006) and assuming that the encounter strips the outer disc layers,
they argue that $\Delta M/M$ should be equal to $\Delta r/r$ for a mass surface density within the disc $\propto r^{-1}$.
With Eq.~(4) from {{Olczak} {et~al.}} (2006) one obtains for the disc size after an encounter:

\begin{align}
   \frac{r_{\mathrm{final}} }{r_{\mathrm{init}} } = &\left(1 -\left( \frac{M_2}{M_2 + 0.5 M_1} \right)^{1.2} \ln{\left[ 2.8 \left( {r_{\mathrm{p}}} \right)^{0.1} \right]} \right. \notag\\
   &\left.\times \exp \left\{ -\sqrt{\frac{M_1}{M_2 + 0.5 M_1}} \left[ \left( {r_{\mathrm{p}}} \right)^{3/2} - 0.5 \right] \right\}\right).\label{eq:mass_loss}
\end{align}

Figure~\ref{fig:cmp_mass_loss} shows that Eq.~\eqref{eq:mass_loss} is also only an upper limit for the disc sizes after encounters.
On the one hand, this can be explained by the fact, that in distant encounters, where no mass is removed from the disc, the disc can shrink by angular momentum removal.
On the other hand, our size definition via the point of steepest surface density makes it possible, that a significant amount of bound mass is located
in the low density regions outside of our defined disc size.
However, this low density is not likely to be detected by observations.

Additionally, both approaches of {{de Juan Ovelar} {et~al.}} (2012) suffer from the fact, that the disc size change is dominated by the redistribution of disc material
(see also Sec.~\ref{sec:discussion}) and not by the truncation of the outer disc material as assumed.

In summary, previous approximations and fit formulae are largely unsuitable to describe the disc size after an encounter
other than for a narrow range for equal-mass encounter partners.

\section{Summary}
\label{sec:summary}

In the dense stellar environments of young clusters, tidal interactions can change the sizes of protoplanetary discs.
Performing N-body simulations of such encounters, we confirm earlier results by {{Kobayashi} \& {Ida}} (2001) that close ($\le 100$\,AU)
encounters between equal-mass solar-type stars lead to the shrinking of initially $100$\,AU-sized discs to $30\text{--}50$\,AU.
Naturally this represents only a special case, in clusters the wide spectrum of encounter partners and periastra has to be taken into account.

In this paper we investigated the disc-size change by encounters for the entire parameter space spanned by mass ratio and periastron distance
typically covered in clusters.
The central result of our extensive numerical parameter study is that the disc size after a prograde,
parabolic encounter is a simple function of the periastron distance $r_{\mathrm{peri}} $ and the mass ratio ${m_{\mathrm{12}}} $ of the two stars of the form
\begin{align}
  r_{\mathrm{final}}  = 0.28 \cdot r_{\mathrm{peri}}  \cdot {m_{\mathrm{12}}} ^{-0.32}. \notag
\end{align}

Prograde, parabolic encounters are the most destructive type of encounters.
Inclined and/or hyperbolic encounters lead to less mass loss and therefore larger disc sizes.
However, the parameter dependencies in these types of encounters span a wide parameter range.
We will investigate this extended parameter range in a follow up study (in preparation).

The disc sizes after a star-disc encounter as obtained with the here presented disc-size definition would not necessarily correspond to the final size of a potentially developing planetary system.
The reasons are that on long time scales (several Myr) viscosity leads to an increase in disc size.
Simultaneously highly eccentric particles probably become recircularised through viscous processes
as they pass the inner parts of the disc while being close to their periastron.

Previous work on disc sizes after encounters was often motivated by the search for the solar birth environment.
Here the sudden density drop in the mass distribution at $30\text{--}50$\,AU is interpreted as the result of a close fly-by of another star
during the formation phase of the solar system.
Considering only encounters between equal mass stars an encounter distance of $100\text{--}150$\,AU was deduced.
However, recent results show that encounters of the early solar system with less or more massive stars
are at least just as likely as with another solar-mass star.
Our results show now, that any parabolic, prograde encounter which fulfils the relation
\begin{align}
  0.28 \cdot r_{\mathrm{peri}}  \cdot {m_{\mathrm{12}}} ^{-0.32} = 30\text{--}50~\mathrm{AU}
\end{align}
can lead to a solar-system size disc.

So far the dependencies of protoplanetary disc sizes on parameters like stellar mass etc. are observationally not well constrained.
With the advent of ALMA this will quickly change.
Thus it will be also possible to determine whether dense stellar environments have a significant influence on the disc sizes
and the forming planetary systems.
The here derived dependencies will be a useful tool to determine the corresponding encounter events.

\bibpunct{(}{)}{;}{a}{}{,} % to follow the A&A style

\Online

\begin{appendix}
\section{Relative disc size changes}

Here we present the full numerical results from our parameter study.
In contrast to the presentation of the results in the main part of the paper,
the values are normalised to the initial disc sizes to allow easier application to arbitrary initial disc sizes.
${r_{\mathrm{p}}}$ is the encounter periastron distance normalised to the initial disc size \mbox{(${r_{\mathrm{p}}} = r_{\mathrm{peri}} /r_{\mathrm{init}} $)}
and ${m_{\mathrm{12}}} $ is the mass of the perturbing star normalised to the mass of the disc hosting star (${m_{\mathrm{12}}}  = m_2/m_1$).
The final disc size is obtained by multiplying the initial disc size with the respective value from the table.
In the lower left part of the table, where no values are given, the disc size change is $\lesssim 1\%$,
in the upper right part, no values are given, since they may not be reliable (see also discussion in main part of the paper).

\begin{table*}

\caption{Disc sizes for whole parameter space. ${r_{\mathrm{p}}}$ is the encounter periastron ratio \mbox{(${r_{\mathrm{p}}} = r_{\mathrm{peri}} /r_{\mathrm{init}} $)}
and ${m_{\mathrm{12}}} $ is the mass of the perturbing star normalised to the mass of the disc hosting star (${m_{\mathrm{12}}}  = m_2/m_1$).
}\label{tab:sizes}
\centering
\begin{tabular}{ r | r r r r r r r r r r r r}
${r_{\mathrm{p}}}$ & \multicolumn{12}{c}{${m_{\mathrm{12}}} $}\\
\hline\hline
    &0.30 & 0.50 & 1.00 & 1.50 & 2.00 & 3.00 & 4.00 & 9.00 & 20.00 & 50.00 & 90.00 & 500.00\\
\hline
0.4 &0.21 &   -  &   -  &   -  &   -  &   -  &   -  &   -  &   -  &   -  &   -  &   -  \\
0.5 &0.23 &   -  &   -  &   -  &   -  &   -  &   -  &   -  &   -  &   -  &   -  &   -  \\
0.6 &0.27 & 0.20 &   -  &   -  &   -  &   -  &   -  &   -  &   -  &   -  &   -  &   -  \\
0.7 &0.32 & 0.24 & 0.20 &   -  &   -  &   -  &   -  &   -  &   -  &   -  &   -  &   -  \\
0.8 &0.34 & 0.28 & 0.23 & 0.20 & 0.20 &   -  &   -  &   -  &   -  &   -  &   -  &   -  \\
0.9 &0.39 & 0.30 & 0.27 & 0.23 & 0.21 & 0.18 &   -  &   -  &   -  &   -  &   -  &   -  \\
1.0 &0.42 & 0.34 & 0.28 & 0.26 & 0.23 & 0.23 & 0.18 & 0.17 &   -  &   -  &   -  &   -  \\
1.5 &0.59 & 0.49 & 0.41 & 0.38 & 0.34 & 0.32 & 0.28 & 0.24 & 0.18 &   -  &   -  &   -  \\
2.0 &0.76 & 0.65 & 0.54 & 0.47 & 0.45 & 0.41 & 0.38 & 0.29 & 0.23 & 0.20 &   -  &   -  \\
2.5 &0.86 & 0.79 & 0.66 & 0.60 & 0.56 & 0.49 & 0.46 & 0.38 & 0.28 & 0.21 &   -  &   -  \\
3.0 &0.94 & 0.90 & 0.79 & 0.70 & 0.65 & 0.58 & 0.54 & 0.43 & 0.35 & 0.24 & 0.20 &   -  \\
3.5 &0.99 & 0.96 & 0.89 & 0.81 & 0.77 & 0.69 & 0.62 & 0.49 & 0.39 & 0.28 & 0.26 &   -  \\
4.0 &1.00 & 0.99 & 0.95 & 0.92 & 0.86 & 0.79 & 0.71 & 0.57 & 0.45 & 0.34 & 0.27 &   -  \\
4.5 &1.00 & 0.99 & 0.99 & 0.97 & 0.94 & 0.86 & 0.81 & 0.64 & 0.49 & 0.38 & 0.32 &   -  \\
5.0 &0.99 & 0.99 & 0.99 & 0.99 & 0.98 & 0.94 & 0.89 & 0.70 & 0.55 & 0.41 & 0.35 & 0.20 \\
5.5 &1.00 & 0.99 & 0.99 & 1.00 & 0.99 & 0.98 & 0.94 & 0.78 & 0.59 & 0.46 & 0.39 & 0.23 \\
6.0 &0.99 & 0.99 & 0.99 & 1.00 & 1.00 & 0.99 & 0.98 & 0.83 & 0.66 & 0.48 & 0.41 & 0.24 \\
6.5 &1.00 & 0.99 & 1.00 & 0.99 & 0.99 & 1.00 & 0.99 & 0.90 & 0.70 & 0.53 & 0.45 & 0.26 \\
7.0 &0.99 & 0.99 & 0.99 & 1.00 & 1.00 & 1.00 & 1.00 & 0.94 & 0.76 & 0.56 & 0.47 & 0.28 \\
7.5 &  -  &   -  &   -  &   -  & 0.99 & 1.00 & 0.99 & 0.97 & 0.81 & 0.60 & 0.51 & 0.30 \\
8.0 &  -  &   -  &   -  &   -  &   -  & 1.00 & 1.00 & 0.99 & 0.85 & 0.65 & 0.54 & 0.30 \\
8.5 &  -  &   -  &   -  &   -  &   -  &   -  & 0.99 & 1.00 & 0.91 & 0.68 & 0.56 & 0.34 \\
9.0 &  -  &   -  &   -  &   -  &   -  &   -  & 1.00 & 1.00 & 0.94 & 0.72 & 0.59 & 0.35 \\
9.5 &  -  &   -  &   -  &   -  &   -  &   -  &   -  & 1.00 & 0.97 & 0.77 & 0.62 & 0.38 \\
10.0 &  -  &   -  &   -  &   -  &   -  &   -  &   -  & 1.00 & 0.98 & 0.80 & 0.67 & 0.40 \\
10.5 &  -  &   -  &   -  &   -  &   -  &   -  &   -  & 1.00 & 0.99 & 0.83 & 0.69 & 0.41 \\
11.0 &  -  &   -  &   -  &   -  &   -  &   -  &   -  & 1.00 & 1.00 & 0.88 & 0.72 & 0.42 \\
11.5 &  -  &   -  &   -  &   -  &   -  &   -  &   -  &   -  & 0.99 & 0.91 & 0.76 & 0.43 \\
12.0 &  -  &   -  &   -  &   -  &   -  &   -  &   -  &   -  & 1.00 & 0.93 & 0.79 & 0.46 \\
12.5 &  -  &   -  &   -  &   -  &   -  &   -  &   -  &   -  &   -  & 0.95 & 0.82 & 0.47 \\
13.0 &  -  &   -  &   -  &   -  &   -  &   -  &   -  &   -  &   -  & 0.97 & 0.85 & 0.49 \\
13.5 &  -  &   -  &   -  &   -  &   -  &   -  &   -  &   -  &   -  & 0.98 & 0.88 & 0.52 \\
14.0 &  -  &   -  &   -  &   -  &   -  &   -  &   -  &   -  &   -  & 0.99 & 0.91 & 0.54 \\
14.5 &  -  &   -  &   -  &   -  &   -  &   -  &   -  &   -  &   -  & 0.99 & 0.94 & 0.56 \\
15.0 &  -  &   -  &   -  &   -  &   -  &   -  &   -  &   -  &   -  & 0.99 & 0.94 & 0.56 \\
15.5 &  -  &   -  &   -  &   -  &   -  &   -  &   -  &   -  &   -  & 0.99 & 0.96 & 0.58 \\
16.0 &  -  &   -  &   -  &   -  &   -  &   -  &   -  &   -  &   -  & 0.99 & 0.98 & 0.59 \\
16.5 &  -  &   -  &   -  &   -  &   -  &   -  &   -  &   -  &   -  & 0.99 & 0.99 & 0.61 \\
17.0 &  -  &   -  &   -  &   -  &   -  &   -  &   -  &   -  &   -  & 0.99 & 0.99 & 0.64 \\
17.5 &  -  &   -  &   -  &   -  &   -  &   -  &   -  &   -  &   -  & 0.99 & 0.99 & 0.66 \\
18.0 &  -  &   -  &   -  &   -  &   -  &   -  &   -  &   -  &   -  & 0.99 & 0.99 & 0.67 \\
18.5 &  -  &   -  &   -  &   -  &   -  &   -  &   -  &   -  &   -  & 0.99 & 0.99 & 0.69 \\
19.0 &  -  &   -  &   -  &   -  &   -  &   -  &   -  &   -  &   -  & 1.00 & 0.99 & 0.70 \\
19.5 &  -  &   -  &   -  &   -  &   -  &   -  &   -  &   -  &   -  & 1.00 & 0.99 & 0.72 \\
20.0 &  -  &   -  &   -  &   -  &   -  &   -  &   -  &   -  &   -  & 1.00 & 0.99 & 0.74 \\
22.0 &  -  &   -  &   -  &   -  &   -  &   -  &   -  &   -  &   -  &   -  & 0.99 & 0.81 \\
24.0 &  -  &   -  &   -  &   -  &   -  &   -  &   -  &   -  &   -  &   -  & 0.99 & 0.88 \\
26.0 &  -  &   -  &   -  &   -  &   -  &   -  &   -  &   -  &   -  &   -  & 0.99 & 0.94 \\
28.0 &  -  &   -  &   -  &   -  &   -  &   -  &   -  &   -  &   -  &   -  & 0.99 & 0.98 \\
30.0 &  -  &   -  &   -  &   -  &   -  &   -  &   -  &   -  &   -  &   -  & 1.00 & 1.00 \\
32.0 &  -  &   -  &   -  &   -  &   -  &   -  &   -  &   -  &   -  &   -  & 0.99 & 1.00 \\
34.0 &  -  &   -  &   -  &   -  &   -  &   -  &   -  &   -  &   -  &   -  &   -  & 1.00 \\
\hline\end{tabular}
\end{table*}

\end{appendix}

\end{document}